\documentclass[fleqn]{article}
\usepackage{amsmath}
\usepackage{amssymb}
\usepackage{a4}
\usepackage{latexsym}
\usepackage{epsfig}
\usepackage{theorem}

{\theorembodyfont{\rm} \newtheorem{definition}{Definition}}
{\theorembodyfont{\rm} \newtheorem{example}{Example}}
{\theorembodyfont{\rm} \newtheorem{theorem}{Theorem}}
{\theorembodyfont{\rm} }
{\theorembodyfont{\rm} \newtheorem{lemma}{Lemma}}
{\theorembodyfont{\rm} }

\newcommand{\Ac}{{\cal A}}
\newcommand{\Ic}{{\cal I}}
\newcommand{\Jc}{{\cal J}}
\newcommand{\Kc}{{\cal K}}
\newcommand{\Sc}{{\cal S}}

\newcommand{\Functions}{{\cal F}}
\newcommand{\N}{{\mathbb N}}

\newcommand{\Params}{\mathfrak{P}} 
\newcommand{\Par}{\mbox{\sl Par}}    
\newcommand{\Terms}{{\cal T}}
\newcommand{\Types}{\mathfrak{T}}
\newcommand{\Variables}{{\cal V}}
\newcommand{\Var}{\mbox{\sf Var}}    
\newcommand{\alphabet}{(\Ac, \#, \leq)}
\newcommand{\concldef}{\hspace*{\fill} $\diamond$}
\newcommand{\conclude}{\hspace*{\fill} $\Box$}
\newcommand{\depth}{\mbox{\sl depth}}    
\newcommand{\df}{\;:=\;}
\newcommand{\dfn}{\stackrel{\mbox{\scriptsize def}}{\Longleftrightarrow}}
\newcommand{\dom}{\mbox{\sl dom}}    
\newcommand{\ineq}{\mbox{\sl ineq}}    
\newcommand{\nf}{\mbox{\sl nf}}    
\newcommand{\nfaux}{\mbox{\sl nf}_{\mbox{\scriptsize aux}}}    
\newcommand{\Inst}{\mbox{\sl Inst}}    
\newcommand{\ota}{\mathfrak{A}} 
    
\newcommand{\true}{\mbox{\tt true}}    
\newcommand{\false}{\mbox{\tt false}}    
\newcommand{\AllParSubst}{\mbox{\sl AllParSubst}}

\newcommand{\wkw}[1]{\mbox{\bf \underline{#1}}}
\newcommand{\signature}[3]{#1_{1} \times \ldots \times #1_{#3} \rightarrow #2}

\begin{document}

\vspace*{3cm}
\begin{center}

{\LARGE Combining Inclusion Polymorphism\\
  \vspace{1ex} and Parametric Polymorphism}

\vspace*{2.5cm}

\begin{tabular}{ll}

 {\Large Sabine Glesner} & {\Large Karl Stroetmann}\\
 & \\
 {\small Institut f\"ur Programmstrukturen} & {\small Siemens AG} \\
 {\small und Datenorganisation}             & {\small ZT SE 4} \\
 {\small Lehrstuhl Prof. Goos}              & \\
 {\small Universit\"at Karlsruhe}           & {\small Otto-Hahn-Ring 6}\\
 {\small Postfach 6980, D 76128 Karlsruhe}  & {\small D-81739 M\"{u}nchen}\\
 {\small Tel.: +49 / 721 - 608 7399}        & {\small Tel.: +49 / 89 - 636 49 555}     \\
 {\small Fax: +49 / 721 - 300 47}           & {\small Fax: +49 / 89 - 636 42 284}\\  
 {\small\tt glesner@ipd.info.uni-karlsruhe.de} &
 {\small\tt Karl.Stroetmann@mchp.siemens.de}\\
\end{tabular}

\vspace*{3cm}
{\Large Abstract}
\ \\
\ \\
\end{center}

  We show that the question whether a term is typable is decidable 
  for type systems combining
  inclusion polymorphism with parametric polymorphism provided the
  type constructors are at most unary.  To prove this result we first
  reduce the typability problem to the problem of sol\-ving a system
  of type inequations. The result is then obtained by showing that the
  solvability of the resulting system of type inequations is decidable.

\newpage

\section{Introduction}
As a common agreement, a flexible type system needs to contain
inclusion as well as parametric polymorphism. Unfortunately, such a
flexibility of the type system causes type inference to become hard or
even undecidable. In this paper, we investigate the problem of 
checking the well-typedness of terms in the presence of 
inclusion polymorphism combined with
parametric polymorphism. We show that in this case typability is
decidable, provided that the type constructors are at most unary.
This result has not been stated before. 

The result of our paper can be used for the design of new type systems
that combine both inclusion polymorphism and parametric 
polymorphism.  Type systems of this kind are of interest for
object-oriented  programming languages. In particular, our result is
applicable for the programming language Java which up to now does not
allow for parametric polymorphism but will probably do so in future
versions \cite{myers_etal:1997,odersky_wadler:1997}. 

Another area where our result is applicable is logic 
programming.  A number of type systems have been designed in this area,
e.~g.~\cite{apt:94b,hill_topor:1992,pfenning:92,yardeni:1992}, but the
systems that  
have been implemented so far either offer no inclusion polymorphism at all
\cite{hill:94a,somogyi:95} or impose stronger restrictions
\cite{beierle:95c} than a type system that would be based on our result.

As it stands, our result cannot be applied to functional
programming languages because these languages allow for the binary
type constructor $\rightarrow$ which takes two types $\sigma$ and
$\tau$ and returns the type $\sigma \rightarrow \tau$ of all functions
mapping $\sigma$ to $\tau$.

In general, Tiuryn and Urzyczyn have shown that type inference for a
type system which combines inclusion polymorphism and parametric
polymorphism is undecidable for second-order types
\cite{tiuryn_urzyczyn:1996}.  On the other
hand, type inference for inclusion polymorphism combined with nullary
type constructors is decidable: \cite{mitchell:84,mitchell:91}
presents an algorithm called MATCH, which solves type inequations in
the case that only inequations between nullary type constructors are
allowed.  Fuh and Mishra \cite{fuh:90} introduce a similar algorithm
to solve the same problem.  The logic programming language {\sc
  Protos-L} \cite{beierle:95b} is based on this type system.

This paper is organized as follows: Section \ref{sec:type-language}
contains a definition of the type language.  In section
\ref{sec:well-typed-terms} we define well-typed terms. Moreover, we
show how the question whether a term is well-typed can be reduced to the problem of solving
a system of type inequations. The solvability of these systems 
is shown to be decidable in section \ref{sec-calc-inequs}.
Section \ref{sec:concl} concludes.

\section{Type Language}
\label{sec:type-language}
In this section we first introduce a language for describing types.
Since types behave in many ways like terms, there is also a notion
of {\em substitution}. This notion is defined in subsection
\ref{subsec:par-subs}. 

\subsection{Types}
\label{sec:types}
Types are constructed from type constructors and type parameters.
The set of type constructors is partially ordered.  This ordering is extended 
to types. 

\begin{definition}[Ordered Type Alphabet]
\label{def:ota}
  An {\em ordered type alphabet} is a tuple $\ota = (\Ac, \#,\leq)$
  such that 
 \begin{enumerate}
 \item {The \em type alphabet} $\cal A$ is a finite set of {\em type
     constructors}.  Elements of $\cal A$ are denoted by $K,L,M,
   \cdots$.
 \item $(\Ac,\leq)$ is a partial order.
 \item $\# : \Ac \rightarrow \N$ is a function assigning an {\em
     arity}, $\# K$, to every type constructor $K\in \Ac$.   \concldef
 \end{enumerate}
\end{definition}

\begin{definition}[Types]
  To define {\em types}, we assume that an ordered type alphabet 
  $\ota = \alphabet$ and a set 
  $\Params = \{ \alpha_i : i \in \N \}$ of {\em type parameters} are given.
  Then the set of {\em types} $\Types = \Types(\ota, \Params)$
  is defined inductively:
  \begin{itemize}
   \item $\alpha \in \Types$ for all $\alpha \in \Params$.
   \item If $K\in \Ac$, $\#K = n$, and $\sigma_i\in \Types$ for all $i=1,\ldots,n$, 
         then $K(\sigma_1,\ldots,\sigma_n)\in \Types$. 
  \end{itemize}
  If $\#K = 0$, then we write $K$ instead of $K()$.
\concldef
\end{definition}
Types are denoted by $\pi$, $\varrho$, $\sigma$, and $\tau$,
while parameters are denoted by $\alpha$ and $\beta$. A {\em monotype}
is a type constructed without type parameters. If $\tau$ is a type,
then $\Par(\tau)$ denotes the set of type parameters in $\tau$.

Next, we extend the relation $\leq$ from  $\ota$ to
the set of types $\Types(\ota, \Params)$. 
\begin{definition}[Subtype Relation]
\label{def:subtype_rel}
  Let $\ota = \alphabet$ be an
  ordered type alphabet and let $\Types(\ota, \Params)$ be the
  set of types constructed from $\ota$. 
  Then the subtype relation on $\Types(\ota, \Params)$
  is defined inductively:
 \begin{enumerate}
 \item If $\alpha\in \Params$, then $\alpha\leq\alpha$.
  \item If $K,L \in \Ac$, $\# K =m$, and $\# L =n$, then
    $K(\sigma_1,\ldots,\sigma_m) \leq L(\tau_1,\ldots,\tau_n)$ holds
    iff $K\leq L$ and  $\sigma_i \leq \tau_i$ for all
    $i=1,\ldots,\min(m,n)$.
\concldef 
 \end{enumerate}
\end{definition}
Without further provisons, $(\Types,\leq)$ is not a partial order.
This is shown by the counter example given next.
\begin{example}
  Assume that $\Ac = \{K_1,K_2,L_1,L_2\}$ where $\#K_i = 0$ and $\#L_i
  = 1$ for $i=1,2$. The ordering $\leq$ on $\Ac$ is defined by the
  following chain of
  inequations: \\[0.2cm]
  \hspace*{1.3cm}
  $L_1 \leq K_1 \leq K_2 \leq L_2$. \\[0.2cm]
  Then we have $L_1(K_2) \leq K_1$ and $K_1 \leq L_2(K_1)$. However,
  $L_1(K_2) \not\leq L_2(K_1)$.
  \concldef
\end{example}
This problem is caused by an incompatibility between
the arity function $\#:\Ac \rightarrow \N$
and the ordering of the type alphabet. 
\begin{definition}[Compatible]
\label{def:compatible}
  Assume a type alphabet $\ota = \alphabet$ is given. 
  Then the arity $\#:\Ac \rightarrow \N$ is {\em compatible} with the
  ordering $\leq$ iff the following condition is satisfied
  for all type constructors $K$, $L$, and $M$: \\[0.2cm]
  \hspace*{1.3cm} 
     $K \leq L \;\wedge\; L \leq M \;\Rightarrow\; \min(\#K,\#M) \leq \#L$.
\concldef
\end{definition}
{\bf Convention}: \quad For the rest of this paper we assume the
following: If an ordered type alphabet $\alphabet$ is
given, then $\#$ is compatible with $\leq$.  

\begin{lemma}
  If $\ota = \alphabet$ is an ordered type alphabet,
  then $\bigl(\Types(\ota, \Params), \leq\bigr)$ is a partial order.
\end{lemma}
{\bf Proof}:
  We need to show that the  relation $\leq$ is reflexive, antisymmetric, and transitive.
  In order to prove the reflexivity, we have to show $\sigma \leq \sigma$ 
  for all types $\sigma$. This is done via a trivial induction on $\sigma$.
  To prove the antisymmetry, assume $\sigma \leq \tau$ and $\tau \leq \sigma$.
  We have to show $\sigma = \tau$.
  The proof proceeds by induction on $\sigma$.
  \begin{enumerate}
  \item $\sigma$ is a parameter $\alpha$.
        Because of $\sigma \leq \tau$ we know that $\tau = \alpha$.
  \item $\sigma = L(\sigma_1,\ldots,\sigma_l)$. Then $\tau = M(\tau_1,\ldots,\tau_m)$
        and we must have $L \leq M$ and $M \leq L$.
        Since $\leq$ is a partial order on $\Ac$, we have $L = M$ and $l=m$.
        Further, we have \\[0.2cm]
       \hspace*{1.3cm} $\sigma_i \leq \tau_i$ \quad for all $i=1,\ldots,l$, \quad and \\[0.2cm]
       \hspace*{1.3cm} $\tau_i \leq \sigma_i$ \quad for all $i=1,\ldots,l$. \\[0.2cm]
       The induction hypothesis yields $\sigma_i = \tau_i$ for all $i=1,\ldots,l$ 
       and then $\sigma = \tau$ is immediate.
  \end{enumerate}

\noindent
 To prove the transitivity, assume that $\varrho, \sigma, \tau \in \Types(\ota, \Params)$ are
 given such that $\varrho \leq \sigma$ and $\sigma \leq \tau$. We need to prove
 $\varrho \leq \tau$. The proof proceeds by induction on $\sigma$.
 \begin{enumerate}
 \item $\sigma$ is a parameter $\alpha$.
       Then $\varrho$ is $\alpha$ and, similarly,
       $\tau$ is $\alpha$. Obviously, $\varrho \leq \tau$.
 \item $\sigma$ is $L(\sigma_1,\ldots,\sigma_l)$.
       Then $\varrho = K(\varrho_1,\ldots,\varrho_k)$ and 
       $\tau = M(\tau_1,\ldots,\tau_m)$. 
       The assumption $\varrho \leq \sigma$ yields $K \leq L$ and \\[0.2cm]
       \hspace*{1.3cm} $\varrho_i \leq \sigma_i$ \quad for all $i=1,\ldots,\min(k,l)$ \\[0.2cm]
       and, similarly, the assumption $\sigma \leq \tau$ yields $L \leq M$ and \\[0.2cm]
       \hspace*{1.3cm} $\sigma_i \leq \tau_i$ \quad for all $i=1,\ldots,\min(l,m)$. \\[0.2cm]
       Since $\leq$ is a partial order on $\Ac$, we have $K\leq M$.
       Further, the induction hypothesis shows that \\[0.2cm]
       \hspace*{1.3cm} $\varrho_i \leq \tau_i$ \quad for all $i=1,\ldots,\min(k,l,m)$. \\[0.2cm]
       Since the arity $\#$ is compatible with $\leq$,
       we have $\min(k,m) \leq l$. Therefore,
       $\min(k,l,m) = \min(k,m)$. But then $\varrho \leq \tau$ is immediate. 
\conclude
 \end{enumerate}

\subsection{Parameter Substitutions}
\label{subsec:par-subs}
Types behave in many ways like terms. Therefore there is also a notion
of {\em substitution}. Since type parameters are substituted rather
than variables, these substitutions are called {\em parameter substitutions}.
Parameter substitutions are denoted by
the capital Greek letters $\Theta$, $\Phi$, and $\Psi$.

\begin{definition}[Parameter Substitution] 
    A {\em parameter substitution} $\Theta$ is a
    finite set of pairs of the form \\[0.2cm]
    \hspace*{1.3cm} 
       $\bigl[ \alpha_1 \mapsto \tau_1, \ldots, \alpha_n \mapsto \tau_n \bigr]$
    \\[0.2cm]
    where $\alpha_1,\ldots,\alpha_n$ are distinct parameters and
    $\tau_1,..,\tau_n$ are types. It is interpreted as a function mapping type parameters 
    to types:
    \\[0.2cm] 
    \hspace*{1.3cm} $\Theta(\alpha) \df \left\{
         \begin{array}{ll}
            \tau_i & \mbox{if}\; \alpha = \alpha_i; \\
            \alpha      & \mbox{otherwise}.
         \end{array}
       \right.
       $ \\[0.2cm]
       This function is extended to types homomorphically: \\[0.2cm]
       \hspace*{1.3cm} $\Theta\bigl(F(\sigma_1,\ldots,\sigma_n)\bigr)
       \df F\bigl(\Theta(\sigma_1),\ldots,\Theta(\sigma_n)\bigr)$.
       \\[0.2cm]
       We use a postfix notation to denote the result of evaluating
       $\Theta$ on a type $\tau$, i.e.~we write $\tau\Theta$ instead
       of $\Theta(\tau)$. 

    The {\em domain} of $\Theta$ 
    is defined as  $\dom(\Theta) \df \{ \alpha \;|\; \alpha \not= \alpha\Theta \}$.
    The set of parameters appearing in the range of a parameter substitution $\Phi$
    is defined as \\[0.2cm]
    \hspace*{1.3cm} 
       $\Par(\Phi) \df \bigcup\{ \Par(\alpha\Phi) \;|\; \alpha \in \dom(\Phi) \bigr\}$. \\[0.2cm]
    A parameter substitution is called a {\em parameter renaming} iff it has the form \\[0.2cm]
    \hspace*{1.3cm} 
       $[ \alpha_1 \mapsto \alpha_{\pi(1)}, \ldots, \alpha_n \mapsto
    \alpha_{\pi(n)} ]$ \\[0.2cm] 
    where $\pi$ is a permutation of the set $\{1,\ldots,n\}$.

    If $\Theta_1$\/ and $\Theta_2$\/ are parameter substitutions, then their {\em composition}
    $\Theta_1 \circ \Theta_2$\/ is defined such that 
    $\alpha(\Theta_1 \circ \Theta_2) = (\alpha\Theta_1)\Theta_2$\/ 
    holds for all type parameters $\alpha$.
\concldef 
\end{definition}

Parameter substitutions respect the ordering $\leq$ on $\Types$.
\begin{lemma}
\label{lemma:par_subst_order}
  If $\Theta$ is a parameter substitution and $\sigma,\tau\in \Types$,
  then \\[0.2cm] 
  \hspace*{1.3cm} $\sigma \leq \tau \;\Rightarrow\; \sigma\Theta \leq
  \tau\Theta$.
\end{lemma}
{\bf Proof:} 
The proof is done by an induction
following the definition of $\sigma \leq \tau$.
\begin{enumerate}
\item The case $\alpha \leq \alpha$ is obvious.
\item If $\sigma = K(\sigma_1, \ldots, \sigma_m) \leq 
      L(\tau_1, \ldots, \tau_n) = \tau$, then $K\leq L$ and
      $\sigma_i \leq \tau_i$ for $i= 1, \ldots, \min(m,n)$. 
      Using the induction hypothesis we have
      $\sigma_i \Theta \leq \tau_i \Theta$ for all relevant $i$. Therefore
      $K(\sigma_1\Theta, \ldots, \sigma_m\Theta) \leq 
       L(\tau_1\Theta, \ldots, \tau_n\Theta)$.
      \concldef 
\end{enumerate}

\section{Well-Typed Terms}
\label{sec:well-typed-terms}
We define the set of well-typed terms in the first subsection. Then in
subsection \ref{subec:type-checking} we reduce the question whether a
term is well-typed to the solvability of a system of type inequations.

\subsection{Definition of Well-Typed Terms}
We assume a set of functions symbols $\Functions$ and a set of variables $\Variables$ to be given.
Every function symbol $f\in\Functions$ is supposed to have an {\em arity}.
\begin{definition}[Terms]
 The set of terms $\Terms(\Functions, \Variables)$ is defined inductively:
 \begin{enumerate}
  \item If $v\in \Variables$, then $v\in \Terms(\Functions, \Variables)$.
  \item If $f\in\Sigma$, $f$ is $n$-ary, and 
        $t_1, \ldots, t_n \in \Terms(\Functions, \Variables)$, 
        then $f(t_1, \ldots, t_n) \in \Terms(\Functions, \Variables)$.
 \end{enumerate}
 The set of variables occurring in a term $t$ is defined by an obvious
 inductive definition and denoted by $\Var(t)$. If this set is empty, then $t$ is
 called a {\em closed term}.
 The set of closed terms is denoted by $\Terms(\Functions)$.
    \concldef
\end{definition}

\begin{definition}[Signature]
\label{def:sig}
  If $f$ is $n$-ary, then its {\em signature} is a string of $n+1$ types.
  If $\sigma_1 \ldots \sigma_n\tau$ is the signature of $f$, then
  this is communicated by writing \\[0.2cm]
  \hspace*{1.3cm} 
    $f: \signature{\sigma}{\tau}{n}$. \\[0.2cm]
  In the following, we assume that every function symbol $f$ has a signature.

  A signature $\varrho_1 \ldots \varrho_n \pi$ is {\em appropriate} 
  for a function symbol $f$ iff \\[0.2cm]
  \hspace*{1.3cm} $f: \signature{\sigma}{\tau}{n}$ \\[0.2cm]
  and there exists a parameter substitution $\Theta$ such that
  $\pi = \tau\Theta$ and $\varrho_i = \sigma_i\Theta$ for $i=1,\ldots,n$.
  \concldef
\end{definition}

\begin{definition}[Type Assignment]
A {\em type annotation} is a pair written as $t: \tau$ where $t$ is a
term and $\tau$ is a type.  The type annotation $t:\tau$ is called a
{\em variable annotation} if $t$ is a variable. 
If $\Gamma = \{ x_1: \tau_1, \ldots, x_n : \tau_n \}$ is a finite set
of variable annotations such that the variables $x_i$ are pairwise distinct, 
then we call $\Gamma$ a {\em type assignment}. 
If $\Gamma = \{ x_1: \tau_1, \ldots, x_n : \tau_n \}$ is a type assignment, then 
we regard $\Gamma$  as a function with domain $\{x_1,\ldots,x_n\}$
 mapping the variables $x_i$ to the types $\tau_i$, i.e.~we have $\Gamma(x_i) = \tau_i$
for $i=1,\ldots,n$ and $\dom(\Gamma) = \{x_1,\ldots,x_n\}$.
\concldef
\end{definition}

\begin{definition}[Well-Typed Term]
\label{def:well-typed-term}
  The notion of a {\em well-typed} term is defined via a binary
  relation $\vdash$ taking as its first argument a type assignment
  and as its second argument a type annotation.  The
  definition of $\vdash$ is done inductively:
 \begin{enumerate}
  \item If $\Gamma(x) \leq \pi$, then \\[0.2cm]
        \hspace*{1.3cm} $\Gamma \vdash x:\pi$.
  \item If we have 
    \begin{enumerate}
    \item $\Gamma \vdash s_i:\varrho_i$ for all $i=1,\ldots,n$,
    \item $\sigma_1 \times \ldots \times \sigma_n \rightarrow \tau$ is
      appropriate for $f$,
    \item $\varrho_i \leq \sigma_i$ for all $i=1,\ldots,n$, and
    \item $\tau \leq \pi$,
    \end{enumerate}
      then \hspace*{1.3cm} $\Gamma \vdash f(s_1,\ldots,s_n):\pi$.
    \end{enumerate}
    A term $t$ is {\em well-typed} iff there exist a type assignment
    $\Gamma$ and a type $\tau$ such that $\Gamma \vdash t:\tau$.  We
    read $\Gamma \vdash t:\tau$ as ``$\Gamma$ {\em entails}
    $t:\tau$''.  We call $\Gamma \vdash t:\tau$ a {\em type
    judgement}. 
\concldef  
\end{definition}

\subsection{Type Checking}
\label{subec:type-checking}
In this subsection, we reduce the question whether a term is well-typed
to the solvability of a
system of type inequations.  Here, a {\em type inequation} is a pair
of types written as $\sigma \preceq \tau$.  A parameter substitution
$\Theta$ {\em solves} a type inequation $\sigma \preceq \tau$ 
(denoted $\Theta \models \sigma \preceq \tau$) if
$\sigma\Theta \leq \tau\Theta$.  A {\em system} of type inequations is
a set of type inequations.  A parameter substitution $\Theta$ {\em
  solves} a system of type inequations $\Ic$ (denoted 
$\Theta \models \Ic)$ iff $\Theta$ solves every
type inequation in $\Ic$.

Assume that $\Gamma$ is a type assignment and $t:\tau$ is a type annotation
such that $\Var(t) \subseteq \dom(\Gamma)$.
We define a function $\ineq(\Gamma,t:\tau)$ by induction on $t$ such that
$\ineq(\Gamma,t:\tau)$ is a system of type inequations. 
A parameter substitution $\Theta$ will solve $\ineq(\Gamma,t:\tau)$ 
iff $\Gamma\Theta \vdash t:\tau\Theta$.  The
inductive definition of $\ineq(\Gamma,t:\tau)$ is given as follows:
\begin{enumerate}
\item $\ineq(\Gamma,x:\tau) \df \{ \Gamma(x) \preceq \tau \}$
\item Assume the signature of $f$ is given as 
      $f:\sigma_1 \times \ldots \times \sigma_n \rightarrow \sigma$,
      where the type parameters have been appropriately renamed so that they are
      {\em new}, i.e.~the new parameters may occur neither in $\Gamma$ nor 
      in $\tau$ nor in any of the signatures used to construct
      $\ineq(\Gamma,s_i:\sigma_i)$ for some $i=1,\dots,n$.
      Then \\[0.2cm]
      \hspace*{1.3cm} 
        $\ineq\bigl(\Gamma,f(s_1,\ldots,s_n):\tau\bigr) \df 
         \{ \sigma \preceq \tau \} \cup 
         \bigcup\limits_{i=1}^n \ineq(\Gamma,s_i:\sigma_i)$.
\end{enumerate}

Before starting with the proofs of the soundness and completeness for
the above transformation, we state some definitions: 
If $\Gamma$ is a type assignment and $t:\tau$ is a type annotation, then
$\Gamma \triangleright t:\tau$ is called a {\em hypothetical type judgement}.
A parameter
substitution $\Theta$ {\em solves} a hypothetical type judgement 
$\Gamma \triangleright t: \tau$ iff $\Gamma \Theta \vdash t:\tau\Theta$ holds. 
A {\em type constraint}
is either a type inequation or a hypothetical type judgement. A parameter
substitution $\Theta$ {\em solves} a set of type constraints $C$ iff it solves
every type inequation and every hypothetical type judgement in $C$. This is written
$\Theta \models C$. We define a rewrite relation on sets of
type constraints. It is the least transitive relation $\leadsto$ such that:
\begin{enumerate}
 \item $C \cup \{ \Gamma \triangleright x: \tau \} \leadsto C \cup \{
   \Gamma(x) \preceq \tau \}$
 \item Assume that the signature of $f$ is given as $f: \sigma_1
   \times \cdots \times \sigma_n \rightarrow \sigma$ where the type
   parameters have been appropriately renamed so that they are new. Then \\[0.2cm]
\hspace*{1.3cm}
   $C\cup \{ \Gamma \triangleright f(s_1, \ldots, s_n) : \tau \} \leadsto C
   \cup \{ \sigma \preceq \tau \} \cup \bigcup_{i=1}^n \{ \Gamma
   \triangleright s_i : \sigma_i \}$.
\end{enumerate}
If a hypothetical type judgement $\Gamma \triangleright t:\tau$ is given, then the two
rewrite rules can be used repeatedly until the set $\ineq(\Gamma,t:\tau)$ 
is derived. This is easily seen by induction on
$t$. Furthermore, the rewrite relation $\leadsto$ satisfies the
following invariants:
\begin{enumerate}
 \item $(\Theta \models C_2) \wedge (C_1 \leadsto C_2) \Rightarrow
   (\Theta \models C_1)$ \hfill{($\mbox{I}_1$)}
 \item $(\Theta \models C_1) \wedge (C_1 \leadsto C_2) \Rightarrow
   \exists \Psi.(\Theta \subseteq \Psi \wedge \Psi \models C_2)$
   \hfill{($\mbox{I}_2$)}
\end{enumerate}
Before proving these invariants, we show that they suffice to verify
the soundness and completeness of our transformation. 

\begin{theorem}[Soundness of the Transformation] \label{soundness}
  Assume $\Gamma$ is a type assignment and $t: \tau$ is a type
  annotation. If $\Theta \models \ineq(\Gamma,t:\tau)$, then
  $\Gamma\Theta \vdash t : \tau\Theta$.
\end{theorem}
{\bf Proof:}
Since the assumption is  $\Theta\models \ineq(\Gamma, t:\tau)$ and we know that
$\{ \Gamma \triangleright t:\tau \} \;\leadsto\; \ineq(\Gamma,t:\tau)$, 
the invariant ($\mbox{I}_1$) shows that
$\Theta \models \{\Gamma \triangleright t: \tau\}$.
By definition, this implies $\Gamma \Theta \vdash t: \tau\Theta$. \conclude

\begin{theorem}[Completeness of the Transformation]
  Assume $\Gamma$ is a type assignment, $t: \tau$ is a type
  annotation, and $\Theta$ is a parameter substitution such that
  $\Gamma\Theta \vdash t: \tau\Theta$. Then, $\Theta$ can be extended
  to a parameter substitution $\Phi$ that is a solution  of $\ineq(\Gamma, t:\tau)$.
\end{theorem}
{\bf Proof:}
$\Gamma\Theta \vdash t:\tau\Theta$ implies 
$\Theta \models \Gamma \triangleright t:\tau$. 
Since $\{\Gamma \triangleright t:\tau \} \;\leadsto\; \ineq(\Gamma, t:\tau)$, the
invariant ($\mbox{I}_2$) shows that $\Theta$ can be extended to a 
parameter substitution $\Phi$ 
such that $\Phi \models \ineq(\Gamma, t: \tau)$. 
 \conclude \\

\noindent {\bf Proof of ($\mbox{I}_1$):} According to the definition
of the rewrite relation $\leadsto$, it suffices to consider the
following two cases:
\begin{enumerate}
 \item $C_1 = C\cup \{ \Gamma \triangleright x:\tau \} \leadsto C\cup \{
   \Gamma(x) \preceq \tau \} = C_2$. The assumption is that $\Theta
   \models C_2$. Then $\Theta \models C$ and $\Gamma(x)\Theta \leq
   \tau\Theta$. Therefore, $\Gamma\Theta \vdash x : \tau \Theta$ 
   showing $\Theta \models C_1$. 
 \item $C_1 = C \cup \{ \Gamma \triangleright f(s_1, \ldots, s_n): \tau \}
   \leadsto C \cup \{ \sigma\preceq \tau \} \cup \bigcup_{i=1}^n \{
   \Gamma \triangleright s_i:\sigma_i \} = C_2$, where 
   $f: \sigma_1 \times \cdots \sigma_n \rightarrow \sigma$.
   According to the assumption, we have $\Theta\models C$,
   $\sigma\Theta \leq \tau\Theta$, and 
   $\Theta \models \Gamma \triangleright s_i :\sigma_i$ for $i=1, \ldots, n$. 
   Then $\Gamma\Theta \vdash s_i : \sigma_i \Theta$ for $i=1, \ldots, n$. 
   Therefore, $\Gamma\Theta \vdash f(s_1, \ldots, s_n): \tau\Theta$ 
   and that yields the claim. \conclude
\end{enumerate}

To prove the invariant ($\mbox{I}_2$) we need the following lemma, which
follows directly from Defs.~\ref{def:sig} and
\ref{def:well-typed-term}. 
\begin{lemma}
\label{lemma-typable-terms}
 Suppose that $t= f(s_1, \ldots, s_n)$ and $f:\sigma_1 \times \cdots
 \times \sigma_n \rightarrow \sigma$ . Then
 $\Gamma \vdash t:\tau$ iff there is a parameter substitution $\Theta$
 such that $\sigma\Theta \leq \tau$ and $\Gamma \vdash s_i :\sigma_i
 \Theta$ for all $i= 1, \ldots, n$. 
\end{lemma}

\noindent {\bf Proof of ($\mbox{I}_2$):} Again, it suffices to
consider the following two cases corresponding to the definition
of the relation $\leadsto$:
\begin{enumerate}
 \item $C_1 = C\cup \{ \Gamma \triangleright x:\tau \} \leadsto C\cup \{
   \Gamma(x) \preceq \tau \} = C_2$. The assumption is that $\Theta
   \models C_1$. Then $\Theta\models C$ and $\Gamma \Theta \vdash x:
   \tau \Theta$. Therefore, $\Gamma(x)\Theta \leq \tau \Theta$. Define
   $\Psi := \Theta$.    
 \item $C_1 = C \cup \{ \Gamma \triangleright f(s_1, \ldots, s_n): \tau \}
   \leadsto C \cup \{ \sigma\preceq \tau \} \cup \bigcup_{i=1}^n \{
   \Gamma \triangleright s_i:\sigma_i \} = C_2$, \\
   where $f: \sigma_1 \times \cdots \sigma_n \rightarrow \sigma$.
   W.l.o.g.~we assume that the type parameters occurring in this
   signature do not occur in $\dom(\Theta)$, since the type parameters
   in the signature can be renamed. According to our assumption, we
   have $\Theta\models C$ and 
   $\Theta \models \{ \Gamma \triangleright f(s_1, \ldots, s_n): \tau\}$. 
   The latter implies $\Gamma\Theta \vdash
   f(s_1,\ldots, s_n): \tau\Theta$. Lemma \ref{lemma-typable-terms}
   shows that there is a parameter substitution $\Phi$ such that
   $\Gamma\Theta \vdash s_i : \sigma_i\Phi$ for all $i=1, \ldots, n$
   and $\sigma\Phi \leq \tau \Theta$. We can assume that $\dom(\Phi)$
   contains only type parameters occurring in the signature of
   $f$. Then $\dom(\Theta) \cap \dom(\Phi) = \emptyset$. Define $\Psi
   := \Theta \cup \Phi$. \conclude
\end{enumerate}

When checking whether a term $t$ is well-typed we want to compute
a type assignment $\Gamma$ and a type $\tau$ such that 
$\Gamma \vdash t:\tau$ holds. To this end,
we define a most general type assignment $\Gamma_{\mathrm{init}}$ and
a most general type $\tau_{\mathrm{init}}$: Let $\Var(t)$ be the
variables in $t$. Define $\Gamma_{\mathrm{init}} = \bigcup_{x\in
  \mathsf{Var}(t)} \{ x: \alpha_x\}$ and $\tau_{\mathrm{init}} = \alpha$ where
$\alpha_x$ and $\alpha$ are distinct new type parameters. 
The claim now is that $t$ is well-typed if and only if the set of type constraints 
$\ineq\bigl( \Gamma_{\mathrm{init}}, t:\tau_{\mathrm{init}} \bigr)$ is solvable.

\noindent
{\bf Proof}: ``$\Rightarrow$'':
Assume $t$ is well-typed. Then there exists a type assignment $\Gamma$ and a type 
$\tau$ such that $\Gamma \vdash t:\tau$.  Define a parameter substitution 
$\Theta$ by setting 
$\Theta(\alpha_x) = \Gamma(x)$ for $x \in \Var(t)$ and
$\Theta(\alpha) = \tau$.  Then we have $\Gamma(x) = \Gamma_{\mathrm{init}}(x)\Theta$
and $\tau = \tau_{\mathrm{init}}\Theta$ and therefore
$\Theta \models \bigl\{ \Gamma_{\mathrm{init}} 
\triangleright t: \tau_{\mathrm{init}} \bigr\}$.
Since \\[0.2cm]
\hspace*{1.3cm}
  $\bigl\{ \Gamma_{\mathrm{init}} \triangleright t: \tau \bigr\} \leadsto
  \ineq\bigl(\Gamma_{\mathrm{init}},t:\tau_{\mathrm{init}}\bigr) $, \\[0.2cm]
the invariant ($\mbox{I}_2$) shows that there exists a parameter
substitution $\Psi$ such that \\[0.2cm]
\hspace*{1.3cm}
$\Psi \models \ineq(\Gamma_{\mathrm{init}},t:\tau_{\mathrm{init}})$. 
\vspace*{0.2cm}

\noindent
``$\Leftarrow$'':
On the other hand, if 
$\Psi \models \ineq\bigl(\Gamma_{\mathrm{init}}, t:\tau_{\mathrm{init}}\bigr)$, 
then Theorem \ref{soundness} shows
that $\Gamma_{\mathrm{init}}\Psi \vdash t:\tau_{\mathrm{init}}\Psi$ holds.  \conclude

Therefore, the problem whether a term $t$ is well-typed
is reduced to the problem of solving systems of type inequations.

\section{Solving Systems of Type Inequations}
\label{sec-calc-inequs}
In this section, we assume that type constructors are at most unary,
i.e., given an ordered type alphabet $\ota = \alphabet$ we have that
$\# K \leq 1$ for all $K \in \Ac$.  We show that then it is decidable
whether a system $\Sc$ of type inequations is solvable.  To this end we
present an algorithm which effectively tests all possible instantiations
for the type parameters in the type
inequations. The fact that the type constructors are at most unary enables
us to guarantee three important properties during this instantiation
process: We do not create any additional parameters; we do not
increase the overall number of inequations; and the depth of the terms
in the type inequations does not increase. Therefore we can generate
only finitely many systems of instantiated type inequations. If one of these
systems is solvable, then we can construct a solution for $\Sc$.

\subsection{Some Definitions}
We start with some definitions necessary to formulate the algorithm for
checking the solvability of systems of type inequations. 
\subsubsection{Solvability and Equivalence of Type Inequations}
A system of type inequations $\Ic$ is
solvable (denoted $\Diamond \Ic$) iff there is a parameter
substitution $\Phi$ such that $\Phi \models \Ic$.  Two type
inequations $I_1$ and $I_2$ are {\em equivalent} (denoted $I_1 \approx
I_2$) iff a parameter substitution $\Phi$
solves $I_1$ if and only if $\Phi$ solves $I_2$: \\[0.2cm]
\hspace*{1.3cm} $I_1 \approx I_2 \dfn
\forall \Phi \cdot ( \Phi \models I_1 \Leftrightarrow \Phi \models I_2 )$ \\[0.2cm]
A type inequation $I$ is equivalent to $\true$ (denoted $I \approx
\true$) iff every parameter substitution solves $I$, it is equivalent
to $\false$ (denoted $I \approx \false$) iff no parameter substitution
solves $I$.  Two systems of type inequations $\Ic_1$ and $\Ic_1$ are
{\em equivalent} (denoted $\Ic_1 \approx \Ic_2$) iff a parameter
substitution $\Phi$ solves $\Ic_1$ if and only if $\Phi$ solves $\Ic_2$: \\[0.2cm]
\hspace*{1.3cm} $\Ic_1 \approx \Ic_2 \dfn
\forall \Phi \cdot ( \Phi \models \Ic_1 \Leftrightarrow \Phi \models \Ic_2 )$ \\[0.2cm]
Next, a system of type inequations $\Ic$ is equivalent to a set of
systems of type inequations $\mathfrak{I}$ (denoted $\Ic \approx
\mathfrak{I}$) iff $\Ic$ is solvable if and only if there
is a system $\Jc \in \mathfrak{I}$ such that $\Jc$ is solvable: \\[0.2cm]
\hspace*{1.3cm} $\Ic \approx \mathfrak{I} \dfn \bigl( \Diamond \Ic
\Leftrightarrow \exists \Jc\in\mathfrak{I} \cdot \Diamond \Jc \bigr)$.
\\[0.2cm]
To proceed, we define the depth of a type inductively:
\begin{enumerate}
\item $\depth(\alpha) \df 0$ for all type parameters $\alpha$.
\item $\depth(K) \df 1$ for all nullary type constructors $K$.
\item $\depth(K(\sigma)) \df 1 + \depth(\sigma)$.
\end{enumerate}
The depth of a type inequation is defined by taking the maximum: \\[0.2cm]
\hspace*{1.3cm} 
  $\depth( \sigma \preceq \tau ) \df \max\bigl( \depth(\sigma), \depth(\tau) \bigr)$. \\[0.2cm]
Furthermore, we define $\depth(\true) := \depth(\false) := 0$.
The function $\depth$ is then extended to systems of type inequations: \\[0.2cm]
\hspace*{1.3cm} 
  $\depth(\Ic) \df \max \bigl\{ \depth(I) \;|\; I \in \Ic \bigr\}$. \\[0.2cm]
The depth of a parameter substitution $\Phi$ is defined as \\[0.2cm]
\hspace*{1.3cm} 
  $\depth(\Phi) \df \max \bigl\{ \depth(\alpha\Phi) \;|\; \alpha \in \dom(\Phi) \bigr\}$. \\[0.2cm]
We define the depth of the empty parameter substitution  as $0$.
A system of inequations $\Ic$ is {\em solvable at depth k} (denoted $\Diamond_k \Ic$) 
iff there is a closed parameter substitution $\Phi$ such that $\Phi \models \Ic$ and
$\depth(\Phi) \leq k$.

\subsubsection{Definition of $\nf$}
The function $\nf$ takes a type inequation as input
and either produces an equivalent type inequation or yields
$\true$ or $\false$. The function is defined inductively.
\begin{enumerate}
\item $\nf(\alpha \preceq \sigma) \df \alpha \preceq \sigma$ and
      $\nf(\sigma \preceq \alpha) \df \sigma \preceq \alpha$ for every type parameter $\alpha$.
\item $\nf( K \preceq L ) \df \left\{
      \begin{array}{ll}
          \true   & \mbox{iff}\; K \leq L \mbox{;} \\
          \false  & \mbox{else.}
      \end{array}
      \right.$
\item $\nf\bigl( K \preceq L(\tau) \bigr) \df \left\{
      \begin{array}{ll}
          \true   & \mbox{iff}\; K \leq L \mbox{;} \\
          \false  & \mbox{else.}
      \end{array}
      \right.$
\item $\nf\bigl( K(\sigma) \preceq L \bigr) \df \left\{
      \begin{array}{ll}
          \true   & \mbox{iff}\; K \leq L \mbox{;} \\
          \false  & \mbox{else.}
      \end{array}
      \right.$
\item $\nf\bigl( K(\sigma) \preceq L(\tau) \bigr) \df \left\{
      \begin{array}{ll}
          \nf(\sigma \preceq \tau) & \mbox{iff}\; K \leq L \mbox{;} \\
          \false  & \mbox{else.}
      \end{array}
      \right. $ \concldef
\end{enumerate}

\noindent
It is easy to see that $\nf(I) \approx I$ holds for every inequation $I$.
We extend the function $\nf$ to systems of type inequations.
First, we define an auxiliary function $\nfaux$: \\[0.2cm]
\hspace*{1.3cm} 
  $\nfaux(\Ic) \df \bigl\{ \nf(I) \,|\, I \in \Ic \wedge \nf(I) \not= \true \bigr\}$.
  \\[0.2cm]
Then, the function $\nf(\Ic)$ is defined as \\[0.2cm]
\hspace*{1.3cm} 
  $\nf(\Ic) \df \left\{
  \begin{array}{ll}
     \{\false\} & \mbox{if}\; \false \in\nfaux(\Ic) \mbox{;} \\
     \nfaux(\Ic) & \mbox{otherwise.}  \\
  \end{array} \right.
  $ \\[0.2cm]
It is easy to see that $\Ic \approx \nf(\Ic)$ for any system of type inequations $\Ic$.

\subsubsection{Definition of $\AllParSubst$}
Next, we define the function $\AllParSubst$. The input to $\AllParSubst$ is a 
finite set $A$ of type parameters. 
The output is the set of parameter substitutions $\Phi$
such that $\dom(\Phi) \subseteq A$, 
$\depth(\Phi) \leq 1$, and $\Par(\alpha\Phi) \subseteq \{ \alpha\}$ but
$\alpha\Phi \not=\alpha$ for all $\alpha\in A$. 
Therefore, $\AllParSubst(A)$ is equal to the set \\[0.2cm]
\hspace*{1.3cm} 
  $\bigl\{ \Phi \;|\; \dom(\Phi) \subseteq A \;\wedge\; 
            \depth(\Phi) \leq 1 \;\wedge\; 
            \forall \alpha\in A \cdot \Par(\alpha\Phi) \subseteq \{\alpha\}
                                \wedge \alpha\Phi \not=\alpha 
  \bigr\}$. 
  \\[0.2cm]
The function $\AllParSubst$ has the following properties:
\begin{enumerate}
\item $\AllParSubst(A)$ is finite.

      This is true because the type alphabet is assumed to be finite. Therefore,
      given a finite set $A$ of type parameter there are only finitely many types 
      $\tau$ such that $\depth(\tau) \leq 1$ and $\Par(\tau) \subseteq A$.
      But then $\AllParSubst(A)$ must be finite, too.

\item If $\Psi$ is a parameter substitution such that $\depth(\Psi) = n \geq 1$
      and $\Par(\Psi) = \emptyset$,
      then there exist parameter substitutions $\Phi_1$ and $\Phi_2$
      such that 
      \begin{enumerate}
      \item $\Phi_1 \in \AllParSubst(\dom(\Psi))$,
      \item $\depth(\Phi_2) = n-1$, and
      \item $\Phi = \Phi_1 \circ \Phi_2$.
      \end{enumerate}
      To prove this, assume 
      $\Psi = [ \alpha_1 \mapsto \tau_1, \ldots, \alpha_n \mapsto \tau_n ]$.
      For those $\tau_i$ such that $\depth(\tau_i) > 1$, we must have 
      $\tau_i = L_i(\sigma_i)$ for some type constructor $L_i$ and some type
      $\sigma_i$ with $\depth(\sigma_i) < \depth(\tau_i)$.
      W.l.o.g.~assume that $\depth(\tau_i) \leq 1$ for all $i=1,\ldots,m-1$
      and $\depth(\tau_i) > 1$ for all $i=m,\ldots,n$.  Then define \\[0.2cm]
      $\Phi_1 := [ \alpha_1 \mapsto \tau_1, \ldots, \alpha_{m-1} \mapsto \tau_{m-1},
                   \alpha_m \mapsto L_m(\alpha_m), \ldots, \alpha_n \mapsto
                   L_n(\alpha_n) ]$ \quad and \\
      $\Phi_2 := [ \alpha_m \mapsto \sigma_m, \ldots, \alpha_n \mapsto \sigma_n ]$. 
      \\[0.2cm] 
      Then the claim is obvious.

\item $\Ic \approx \bigl\{ \Ic\Phi \;|\; \Phi \in \AllParSubst\bigl(\Par(\Ic)\bigr) \bigr\}$.

      Assume $\Phi \models \Ic$ where w.l.o.g.~$\dom(\Phi) \subseteq \Par(\Ic)$.
      Then the previous property shows that
      $\Phi$ can be written as $\Phi_1 \circ \Phi_2$ where 
      $\Phi_1 \in \AllParSubst\bigl(\Par(\Ic)\bigr)$.  But then $\Phi_2 \models \Ic\Phi_1$.

      Conversely, if $\Psi \models \Ic\Phi$ for a substitution 
      $\Phi \in \AllParSubst\bigl(\Par(\Ic)\bigr)$, then $\Phi\circ\Psi \models \Ic$.

\item If $\Phi \in \AllParSubst\bigl(\Par(\Ic)\bigr)$, then
      $\depth\bigr(\nf(\Ic\Phi)\bigl) \leq \depth(\Ic)$.

      Assume $\sigma \preceq \tau$ is an inequation in $\Ic$ of maximal depth.
      First, assume
      $\sigma = K(\sigma')$  and $\tau = L(\tau')$. 
      When going from $\Ic$ to $\nf(\Ic\Phi)$ this inequation either disappears
      or it has the form $\nf(\sigma'\Phi\preceq \tau'\Phi)$. But the depth of this
      inequation is not greater than the depth of the original inequation.

      Next, $\sigma = \alpha$ for a parameter $\alpha$ and $\tau = L(\tau')$. 
      But then $\sigma\Phi$ must have either of the forms $K$ or $K(\sigma')$.
      When going from $\Ic$ to $\nf(\Ic\Phi)$ the inequation $\sigma \preceq \tau$
       either disappears
      or it has the form $\nf(\sigma'\Phi\preceq \tau'\Phi)$. Again the depth of this
      inequation is not greater than the depth of the original inequation.
      The remaining cases are similar.
\end{enumerate}

\subsubsection{Definition of $\Inst$}
 The function $\Inst$ transforms a single system of type inequations
into an equivalent set of systems of type inequations. 
It is defined as \\[0.2cm]
\hspace*{1.3cm} 
  $\Inst(\Ic) \df \bigl\{ \nf(\Ic\Phi) \;|\; \Phi \in \AllParSubst\bigl(\Par(\Ic)\bigr) 
                 \wedge \nf(\Ic\Phi) \neq \{\false\} \bigr\}$. 
                 \\[0.2cm]
The function $\Inst$ has the following properties: 
\begin{enumerate}
\item $\Inst(\Ic)$ is finite.
\item $\Ic \approx \Inst(\Ic)$.
\item If $\Diamond_k \Ic$ and $k \geq 1$, then there is a 
      $\Jc \in \Inst(\Ic)$ such that $\Diamond_{k-1} \Jc$.
\item If $\Diamond_k \Jc$ and  $\Jc \in \Inst(\Ic)$, then 
      $\Diamond_{k+1} \Ic$.
\item If $\Jc \in \Inst(\Ic)$, then $\Par(\Jc) \subseteq \Par(\Ic)$.
\item If $\Jc \in \Inst(\Ic)$, then  $\depth(\Jc) \leq \depth(\Ic)$.
\end{enumerate}
These properties are immediate consequences of the definition of $\Inst$ and the properties
of the function $\AllParSubst$.

\subsection{Deciding Type Inequations}
We present an algorithm for solving (or refuting) systems of type inequations.
The algorithm maintains two sets of systems of inequations. Call theses sets
$\mathfrak{M}$ and $\mathfrak{A}$. 
$\mathfrak{M}$ serves as a memory of systems of type inequations that have already been
encountered, while $\mathfrak{A}$ contains systems of type inequations that can be derived
from $\Ic$ by application of the function $\Inst$.
The algorithm initializes both $\mathfrak{M}$ and $\mathfrak{A}$ to the singleton $\{ \Ic \}$,
where $\Ic$ is the system of type inequations that is to be solved.
After this initialization, the algorithm enters a loop.
In this loop, we compute $\Inst(\Jc)$ for all $\Jc \in \mathfrak{A}$.
Then, we update $\mathfrak{A}$ as follows: \\[0.2cm]
\hspace*{1.3cm} 
  $\mathfrak{A} \df \bigcup \bigl\{ \Inst(\Jc) \;|\; \Jc \in \mathfrak{A} \bigr\} - \mathfrak{M}$
  \\[0.2cm]
that is, we apply $\Inst$ to all systems in $\mathfrak{A}$ and 
we discard those systems that appear already in the memory $\mathfrak{M}$.
If $\emptyset \in \mathfrak{A}$, then 
$\Ic$ is solvable and the algorithm halts with success.
If $\mathfrak{A}$ becomes empty, the algorithm halts with failure.
Otherwise, we update $\mathfrak{M}$ as  \\[0.2cm]
\hspace*{1.3cm} 
  $\mathfrak{M} \df \mathfrak{M} \cup \mathfrak{A}$ \\[0.2cm]
and reenter the loop.
Figure \ref{figure1} specifies the algorithm formally.

\begin{figure}[htbp]
\begin{center}
\leavevmode
\fbox{
\hspace*{0.5cm}
\begin{minipage}{10cm}

\noindent
\wkw{Input}:\quad $\Ic$ \hspace*{\fill} \% system of type inequations to be solved \\
$\mathfrak{M} \df \{\Ic\}$; \\
$\mathfrak{A}_0 \df \{\Ic\}$; \\
$n \df 0$; \\
{\sl Loop}: \\
\hspace*{1cm}
  $\mathfrak{A}_{n+1} \df \bigcup \bigl\{ \Inst(\Jc) \;|\; \Jc \in \mathfrak{A}_n \bigr\} - \mathfrak{M}$
  \\
\hspace*{1cm}
  \wkw{if} $\emptyset \in \mathfrak{A}_{n+1}$ \wkw{then} \\
\hspace*{2cm}
  \wkw{return} \true;  \\
\hspace*{1cm}
  \wkw{end-if};  \\
\hspace*{1cm}
  \wkw{if} $\mathfrak{A}_{n+1} = \emptyset$ \wkw{then} \\
\hspace*{2cm}
  \wkw{return} \false;  \\
\hspace*{1cm}
  \wkw{end-if}; \\
\hspace*{1cm}
  $\mathfrak{M} \df \mathfrak{M} \cup \mathfrak{A}_{n+1}$; \\
\hspace*{1cm}
  $n \df n+1$; \\
\hspace*{1cm}
  \wkw{goto} {\sl Loop}; \\
\end{minipage}
}
\caption{An algorithm for deciding solvability of type inequations. \label{figure1}}
\end{center}
\end{figure}

\begin{lemma}[Termination]
  The algorithm given in Figure \ref{figure1} terminates.
\end{lemma}
{\bf Proof}: 
For every system of inequations $\Jc \in \mathfrak{M}$
the number of inequations in $\Jc$ is less or equal than the number of inequations
in $\Ic$, 
 $\Par(\Jc) \subseteq \Par(\Ic)$, and
 $\depth(\Jc) \leq \depth(\Ic)$. 
Since the type alphabet is finite, the size of $\mathfrak{M}$ must therefore
be bounded. 

Now assume the algorithm given in Figure \ref{figure1} does not terminate.
Then the set $\mathfrak{A}_{n+1}$ can never be empty. Therefore,
every time the loop is executed, the statement $\mathfrak{M} := \mathfrak{M} \cup \mathfrak{A}_{n+1}$
increases the number of elements of the set $\mathfrak{M}$.
But then the size of $\mathfrak{M}$ would increase beyond every bound.
\conclude
\vspace*{0.2cm}

\begin{lemma}[Soundness] \label{sound}
  Assume $n,k\in\N$, $\Jc \in \mathfrak{A_n}$ and $\Diamond_{k} \Jc$.
  Then $\Diamond_{k+n} \Ic$.
\end{lemma}
{\bf Proof}: The proof is given by induction on $n$.
\begin{enumerate}
\item $n=0$: 
      Since $\mathfrak{A}_0 = \{ \Ic \}$ we must have $\Jc = \Ic$ and the claim
      is trivial.
\item $n \rightarrow n+1$: 
      Assume $\Jc \in \mathfrak{A}_{n+1}$ with $\Diamond_{k} \Jc$.
      Then there is a $\Kc \in \mathfrak{A}_n$ such that $\Jc \in \Inst(\Kc)$.
      This implies $\Diamond_{k+1} \Kc$. By i.h.~we have $\Diamond_{(k+1)+n} \Ic$.
      \conclude
\end{enumerate}

\begin{lemma} \label{complete}
 Assume that $\Diamond_k \Ic$ and $k$ is minimal with this property.
 Then for all $n \leq k$ there is a $\Jc \in \mathfrak{A_n}$
 such that $\Diamond_{k-n} \Jc$.
\end{lemma}
{\bf Proof}: The proof is done by induction on $n$.
\begin{enumerate}
\item $n=0$: Obvious.
\item $n \rightarrow n+1$: 
      Assume $\Diamond_k \Ic$ and that $k$ is minimal with this property.
      By i.h.~there is a $\Jc\in\mathfrak{A}_n$ such that $\Diamond_{k-n} \Jc$.
      Then there is a $\Kc \in \Inst(\Jc)$ such that $\Diamond_{k-n-1} \Kc$.
      Assume $\Kc \in \mathfrak{M}$. Since \\[0.2cm]
      \hspace*{1.3cm} 
        $\mathfrak{M} = \bigcup\limits_{i=1}^n \mathfrak{A}_i$, \\[0.2cm]
      there is an $i\leq n$ such that
      $\Kc \in \mathfrak{A}_i$. Therefore Lemma \ref{sound} shows
      $\Diamond_{k-n-1+i}\, \Ic$. Since $k-n-1+i < k$ this contradicts the minimality of
      $k$. This shows that the assumption $\Kc \in \mathfrak{M}$ is wrong and we have 
      $\Kc \in \mathfrak{A}_{n+1}$.
      Because of $\Diamond_{k-(n+1)} \Kc$  the proof is complete.
      \conclude
\end{enumerate}

\begin{theorem}
  The algorithm given in Figure \ref{figure1} is correct.
\end{theorem}
{\bf Proof}:
Assume that $\Ic$ is solvable. Then $\Diamond_n \Ic$ for some $n\in \N$.
By Lemma \ref{complete} we find a $\Jc \in \mathfrak{A}_n$ such that $\Diamond_0 \Jc$
holds. But then $\Jc = \emptyset$ and the algorithm returns {\tt true}.

Assume now that $\Ic$ is not solvable. If the algorithm would return {\tt true},
then $\emptyset \in \mathfrak{A}_n$ for some $n\in \N$. Since $\Diamond_0 \emptyset$
Lemma \ref{sound}  would then give
$\Diamond_n \Ic$. Therefore the algorithm cannot return {\tt true}. Since it
terminates, it must return {\tt false}.
\conclude

\section{Conclusion}
\label{sec:concl}
In this paper we have presented a type system that supports both inclusion
polymorphism and parametric polymorphism.  We were able to prove that for
this type system typability is decidable, provided we use at most unary 
type constructors.  In practice, many interesting type constructors are either nullary
or unary.  Unary type constructors occur naturally when dealing with
container types, e.~g.~types that are interpreted as sets, lists, or
bags.  It is convenient to be able to cast, for example, lists to sets.
This cannot be done with the type system proposed by Mitchell
\cite{mitchell:84}, but is possible with the type system introduced in this paper.

Previously, it has been know that type inference is decidable for a system
that restricts inclusion polymorphism to nullary type constructors
\cite{fuh:90,mitchell:84,mitchell:91}.  On the negative side, Tiuryn and
Urzyczyn \cite{tiuryn_urzyczyn:1996} have shown that the type inference
problem for second-order types is undecidable.  We have shown in this
paper, that typability is decidable for type systems with at most
unary type constructors. 
It is still an open question whether typability is decidable in the case of binary type
constructors.
\vspace*{0.2cm}

\noindent
{\bf Acknowledgement}: The authors would like to thank Pawel Urzyczyn for pointing out 
some technical weaknesses in an earlier version of this paper.

\end{document}